\documentclass{elsart}

\usepackage{harvard}

 \usepackage{graphics}
\usepackage{graphicx}
\usepackage{epsfig}

\usepackage{amssymb}

\def\url#1{{\ttfamily\def\/{/\discretionary{}{}{}}#1}}

\begin{document}

\begin{frontmatter}
\title{Cluster temperatures and non-extensive thermo-statistics}
\author{Steen H. Hansen}
\address{University of Zurich, Winterthurerstrasse 190,
8057 Zurich, Switzerland}

\begin{abstract}
We propose a novel component to the understanding of the temperature
structure of galaxy clusters which does not rely on any heating or
cooling mechanism. The new ingredient is the use of non-extensive
thermo-statistics which is based on the natural generalization of
entropy for systems with long-range interactions.  Such interactions
include gravity and attraction or repulsion due to charges.  We
explain that there is growing theoretical indications for the need of
this generalization for large cosmological structures.  The observed
pseudo temperature is generally {\em different} from the true
thermodynamic temperature, and we clarify the connection between the
two.  We explain that this distinction is most important in the
central part of the cluster where the density profile is most shallow.
We show that the observed pseudo temperature may differ up to a factor
2/5 from the true thermodynamic temperature, either larger or smaller.
In general the M-T and L-T relations will be affected, and the central
DM slope derived through hydrostatic equilibrium may be either more
shallow or steeper.  We show how the true temperature can be extracted
correctly either from the spectrum or from the shape of the Doppler
broadening of spectral lines.\end{abstract}

\end{frontmatter}


\section{Introduction}

Galaxy clusters have been known and studied for many years, and the
radial dependence of cluster temperatures is becoming a testing ground
for models of structure formation and for our understanding of gas
dynamics.  The emerging temperature profile is one where the
temperature increases from the centre to some maximum temperature, and
then decreases again for larger radii.

The central decrement has been much discussed and the possibility of
cooling flows has been explained in excellent reviews, see
\cite{fabian94,donahue03} for references.  From an observational point
of view this central temperature decrement is very well
established~\cite{allen01,peterson03,kaastra03}, and numerical
simulations are now beginning to see it too~\cite{motl03}.  The outer
temperature decrease is well established both
observationally~\cite{markevitch98,kaastra03} and
numerically~\cite{loken02,komatsu01}.

The expected cooling flows are not observed in galaxy clusters, and
numerous explanations have been proposed including various additional
heating or lack-of-cooling mechanisms, see
e.g. \cite{peterson03,donahue03} for references.

We will here propose a new element in the understanding of cluster
temperatures. Our solution has galaxy clusters in {\em kinetic
equilibrium}, but with temperatures defined through non-extensive
thermo-statistics which is the natural generalization of normal
statistical mechanics. The need for non-extensive entropy arises when
particle interactions are not point-like, and includes gravity and
attraction/repulsion due to charges. These are exactly the conditions
for electrons and protons in galaxy clusters, as we will explain in
section~\ref{sec:pro}.

We will structure the paper as follows. First we will consider the
theoretical basis for statistical mechanics with non-extensive
entropy. To clarify the signatures of non-extensive statistics, we
will apply the results to the Coma cluster. This is followed by a
discussion of reasons for using Tsallis statistics, potential problems
and implications of our findings. Finally we offer our conclusions.

\section{Tsallis statistics}
Let us first consider the theoretical basis for statistical mechanics
with non-extensive entropy.  Statistical mechanics for classical gases
can be derived from the Boltzmann-Gibbs assumption for the entropy,
$S_{BG} = - k \sum p_i \cdot {\rm ln}p_i$, where $p_i$ is the
probability for a given particle to be in the state $i$, and the sum
is over all states. For normal gases the probability, $p(v)$,
coincides with the velocity distribution function, $f(v)$.  This
classical statistics can be generalized to Tsallis (also called non-extensive)
statistics~\cite{tsallis88}, which depends on the entropic index $q$
\begin{equation}
S_q = - k \sum_i p_i^q \cdot {\rm ln}_q p_i \, .
\label{stsallis}
\end{equation}
Here the q-logarithm is defined by, ln$_q p = (p^{1-q}-1)/(1-q)$, and
for $q=1$ the normal Boltzmann-Gibbs entropy is recovered, $S_{BG} =
S_1$.  The probabilities still obey, $\sum p_i =1$, while the particle
distribution function is now given by $f(v) = p^q(v)$. Thus, for $q<1$
one privileges rare events, whereas $q>1$ privileges common events.
For a summary of applications see \cite{tsallis99}, and for up to date
list of references see {\tt http://tsallis.cat.cbpf.br/biblio.htm}.

Average values are calculated through the particle distribution
function, and one e.g. has the mean energy~\cite{tsallisMP98}
\begin{equation}
U_q = \frac{\sum p_i^q E_i}{c_p} \, ,
\end{equation}
where $c_p=\sum p_i^q$, and $E_i$ are the energy eigenvalues.
Optimization of the entropy in eq.~(\ref{stsallis}) under the
constraints leads to the probability~\cite{silva98,tsallis03}
\begin{equation}
p_i = \frac{\left[1 - (1-q)\beta_q \left(E_i-U_q\right)  
\right] ^{1/(1-q)}}{Z_q} \, ,
\label{eq:pi1}
\end{equation}
where $Z_q$ normalizes the probabilities, $\beta_q = \beta/c_p$, and
$\beta$ is the optimization Lagrange multiplier associated with the
average energy.  Adding a constant energy, $\epsilon_0$, to all the
energy eigenvalues leads to $U_q \rightarrow U_q + \epsilon_0$, which
leaves all the probabilities, $p_i$, invariant~\cite{tsallisMP98}.  By
defining
\begin{equation}
\alpha = 1 +  (1-q)\beta_q U_q \, ,
\label{eq:alpha}
\end{equation}
eq.~(\ref{eq:pi1}) can be written as
\begin{equation}
p_i = \frac{
\left( 1 - (1-q) (\beta_q/\alpha) \,  E_i
\right) ^{1/(1-q)}}{Z_q'} \, ,
\label{eq:pi2}
\end{equation}
and we see that the probabilities have the shape of 
q-exponential functions
\begin{equation}
p_i = \frac{{\rm exp}_q\left(-\beta_q' \,E_i \right)}{Z_q'} \, ,
\label{eq:pi3}
\end{equation}
where 
\begin{equation}
\beta_q ' = \frac{\beta_q}{\alpha} \, .
\label{eq:beta}
\end{equation}
For $q=1$ one recovers the standard Maxwell distribution with 
$p_i \sim {\rm exp}(-\beta E_i)$.

We now note a very important detail, namely that the distribution
function contains the 'pseudo-inverse temperature', $\beta_q '$, which
differs from the real temperature $\beta_q = \partial S_q/\partial
U_q$, as described in eq.~(\ref{eq:beta}).  If one was blindly to fit
a given spectrum with the shape of eq.~(\ref{eq:pi2}), then one gets
the observed $\beta_q'$, which is not the true (inverse)
temperature. Instead, the true (inverse) temperature is $\beta_q =
\beta_q' \cdot \alpha$ (for a discussion on temperature in
non-extensive statistics see e.g.  \cite{rama,martinez}).

In section~\ref{sec:pro} we will discuss in more detail why and where
one should expect Tsallis statistics to be important for cluster
physics.

A brief summary of our findings so far is the following. For particles
with long-range interactions (e.g. charged particles or particles with
gravitational interactions only) the velocity distribution function
may in general be different from the normal Maxwell distribution,
$f(v) \sim {\rm exp}(-\beta E)$, and will instead follow the shape
given in eq.~(\ref{eq:pi2}) with $f(v)=p^q(v)$, for systems in kinetic
equilibrium. It is important to keep in mind that these distributions
are in kinetic equilibrium, and they are therefore stable even though
the equilibration timescale for electrons and protons is much smaller
than the age of the cluster.  The observed pseudo temperature is
generally different from the true thermodynamic temperature, and
differs by the value of $\alpha$ in eq.~(\ref{eq:alpha}).

\section{An example: Coma}
\label{sec:coma}
To show the signatures of non-extensive statistics, we now wish to
consider observed X-ray data of a cluster structure, and we consider
the central part of NGC 4874, which is near the center of the Coma
cluster~\cite{arnaud}.  This is just to exemplify how one can extract
the true cluster temperature.  In figure 1 we plot the continuum part
of the spectrum for two radial bins near the center of NGC 4874,
namely the center-most bin, where the temperature was observed to drop
to approximately $6.6$ keV, and also the third bin, where the
temperature plateau is already reached with $T\approx 8.5$
keV~\cite{arnaud}.  By 'continuum part' we simply consider the energy
ranges (in keV) $0.2-0.5$, $3.5-3.7$, $4.25-6.3$, $7.2-7.5$, and
$8.7-10$. The data is binned to have signal to noise ratio of 10
sigma, and other binnings leave the spectral shape intact. The
horizontal error-bars represent the bins in energy.  The plotted data
is background subtracted and convolved with the instrument response
matrix.  For this example we are going to assume (incorrectly) that
the response matrix is unity, and can therefore compare theoretical
curves directly with the data-points.  We emphasize that this is only
an example, since the detector response drops at high energies.  As is
clear from the (green) triangles on the figure, the outer radial bin
with particles in the temperature plateau are well fitted with a
normal exponential, corresponding to $q=1$, that is, the continuum is
well fitted with a single exponential (straight dashed line) which
implies that the normal temperature concept is correct.  This is
contrasted with the inner radial bin (red circles), where an
exponential (straight dashed line) gives a rather bad fit to the high
energy part of the spectrum.  Instead, fitting with a shape of
eq.~(\ref{eq:pi2}) leads to agreement, with $q=0.87$ (solid
line). Such value of $q$ corresponds to a true temperature which is
about $20\%$ smaller than the observed temperature if we adopt
$U_q\beta_q=3/2$ (see below) and use eqs.~(\ref{eq:alpha},
\ref{eq:beta}). We emphasize that this is just an example, and in
order to actually extract the true temperature one must convolve with
the instrument response matrix.  For a given cluster region the
theoretical curve could bend either up or down depending on $q$ being
smaller or larger than unity.  There is still no observational
evidence for the necessity of using Tsallis statistics for galaxy
clusters, however, in section~\ref{sec:pro} we will explain that from
a theoretical point of view there is growing evidence that exactly
Tsallis statistics is the correct statistics to consider for large
cosmological structures, and in particular in the central region.

\begin{figure}
\begin{center}
\epsfxsize=8.6cm
\epsfysize=6.4cm
\epsffile{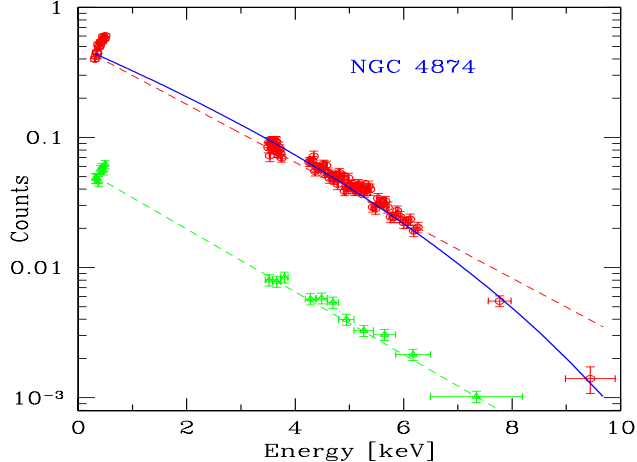}
\end{center}
\caption{The continuum part of the spectra for two radial bins near
NGC 4874. The (green) triangles from the non-central region is well
fitted with an exponential, corresponding to $q=1$.  The (red) circles
are not well fitted with an exponential in the high energy
part. Instead, using formula~(\ref{eq:pi2}) we find agreement when
using $q=0.87$. This corresponds to the true temperature being
overestimated by approximately $20\%$. This figures serves only as an
illustration of the signatures of non-extensive statistics, and the
fitted $q=0.87$ serves only as an example since the detector response
is not included.}
\label{fig:coma}
\end{figure}

One should be very careful when trying to directly compare to previous
temperature measurements. First of all the cluster temperatures are
normally found by a full spectral analysis, and not trivially with
fit-by-eye of the (approximately) exponential tail. The exponential
tail of the photon counts does come from the exponential tail of the
distribution of the plasma particles~\cite{rybicki}, and for a
generalized q-exponential the photon count will follow the generalized
shape of eq.~(\ref{eq:pi3}).  The same statement holds true for
non-thermal bremsstrahlung, and also in our case must the Gaunt
factors be recalculated.  Also the spectral lines are expected to be
different in non-extensive thermo-statistics~\cite{silva98}. A proper
implementation and comparison is beyond the scope of this short
article, so we will leave that for a future analysis.  From figure 1
we see that the non-extensive curve (solid line) is slightly above the
exponential continuum curve for low energies, $E<5$ keV, and it will
be interesting to study in detail if this is connected to the observed
soft excess.

\section{Reasons for using Tsallis statistics}
\label{sec:pro}
An important question to ask is naturally if there is any reason for
considering the non-extensive thermo-statistics, that is, why fit the
continuum with eq.~(\ref{eq:pi2}) and not just with a simple exponential? 
We will now explain that the answer to this question, from a theoretical
point of view, is a definite yes.

Systems with particles having point interaction generally have
extensive entropy, however, systems with long-range interactions do
not have extensive entropy, e.g. dark matter experiences long-range
gravitational interaction (only), and do therefore not obey the rule
of extensivity. 

The electrons and protons in clusters of galaxies experience two forces,
namely the attraction and repulsion from charges when the particles
are nearby each other, and the gravitational attraction when the
particles are charge screened. Both these forces are long-range, and we
will consider them in turn.

Charged particles have long-range repulsion and attraction, and it is
known that a pure electron plasma in 2 dimensions is correctly
described by Tsallis statistics and has entropic index
$q=1/2$~\cite{huang,boghosian96}.  Laboratory experiments of inverse
bremsstrahlung absorption in plasma certainly exhibit non-Maxwellian
form~\cite{liu}, and are infact well described by the generalized form
in eq.~(\ref{eq:pi2})~\cite{tsallissouza}.  Generally, the Tsallis
statistics applies very well to a variety of astrophysical plasma
environments, see e.g. \cite{leubner} for references.   One
should therefore expect that charged electrons and protons, as in galaxy
clusters, may have entropic index different from unity when the
interactions are purely due to the charges, i.e. for distances
shorter than the distances on which the charges are neutralized.

On larger distances the only force is the gravitational attraction.
It is well-known that the velocity distribution for particles in an
equilibrated structure with only gravitational interactions do not
follow a Maxwell distribution. Infact, for the simplest structures,
polytropes, the distribution function is exactly the Tsallis distribution
of eq.~(\ref{eq:pi3})~\cite{plastino}, and not just any random 
non-thermal distribution function
\footnote{The Tsallis shape of the distribution function has also been
seen directly from cosmological N-body simulations, and the $\alpha$
from eq.~(\ref{eq:alpha}) may infact be crucial for the understanding
of the observed relation between the anisotropy and the density slope
of structures of purely gravitating
particles~\cite{hansenmoore}.}. Infact, a simple connection between
the density slope, $ \gamma = d {\rm ln} \rho/d {\rm ln} r$ and the
entropic index, $q$, can be found~\cite{hansen04}, $q =
(10+3\gamma)/(6 + \gamma)$. Here one sees directly that structures
with density slope of $\rho \sim r^{-2}$ have $q\approx 1$, and hence
normal extensive thermo-statistics provides a correct
description. Hence, when the density slope is about $-2$ (as in the
outer part of a beta-model), then normal thermo-statistics applies.
Further, if the particles are not fully ionized then particle
interactions are again point like, and hence one should expect normal
extensive thermodynamics to give a correct description in the outer
cluster region.

However, the inner part of a cluster may have a density slope more
shallow than $-2$, and one should therefore expect that the entropic
index should differ from unity in the central cluster region.

One may wonder if centrally disturbing objects (like a cD galaxy) may
be related to a different entropic index. Observationally the very
central region near the cD galaxy M87 is known to not follow a single
phase structure locally~\cite{matsushita}.
This could come about since
the timescale for reaching kinetic equilibrium (correctly described by
non-extensive thermodynamics) is much shorter than the timescale for
reaching thermal equilibrium.  The thermalization of energetic
particles in a cold gas was infact shown
numerically~\cite{waldeer} to lead to power-law distributions, which
eq.~(\ref{eq:pi2}) indeed is~\cite{tsallis99}.  This question of
timescales can be addressed by considering the spectra from shocks of
merging clusters which are still dynamically young.  If our
interpretation with rapidly achieved kinetic equilibrium is correct,
then such dynamically young structures may have entropic indices
different from unity.  Another approach would be a
numerical study of the equilibration of electron-proton plasma in an
external gravitational field, however, that is a numerically very hard
challenge.

To briefly summarize this section, then there are growing theoretical 
indications that the central region of clusters very well may have
an entropic index different from unity. Since the forces involved
are gravity and electromagnetic attraction and repulsion, which both
are known (experimentally, numerically and theoretically) to have
distribution functions of the Tsallis type, then it is very likely
that the electron distribution infact does follow the shape in 
eq.~(\ref{eq:pi3}).

\section{Implications}
\label{sec:imp}

We saw above that the observed pseudo temperature, $\beta_q'$, differs
from the true thermodynamic temperature, $\beta_q$. How big can this
difference be? It is known~\cite{LSP} that $q$ is bounded from below
by $q>0$, and if we adopt $U_q\beta_q=3/2$ \cite{boghosian,hansen04}, 
then we see that $\alpha$ is bounded from
above by $\alpha=5/2$, and hence the true temperature may be a factor
5/2 smaller than the observed temperature.  Similarly, to assure
positiveness of thermal conductivity Boghosian~\cite{boghosian}
found for an ideal gas that $q$ is bounded from above by $q<7/5$,
which implies a lower bound of $\alpha> 2/5$.
E.g. cooling flow clusters have been observed to have
a central temperature which is at most a factor 3 below the the peak
temperature~\cite{kaastra03,peterson03}, however, 
the true central temperature may thus be different by a factor of 2/5
from the observed value. This may be either larger or smaller depending
on the value of $q$, and may thus reduce or increase the cooling flow
problem.

If our proposed component to the understanding of cluster temperatures
indeed turns out to be correct, then it will have numerous
implications. All quantities that depend on the velocity distribution
function of electrons will be different, this includes in particular
the cooling function, heat conduction coefficient, Gaunt factors, and
the Sunyaev-Zeldovich effect.  The temperature of the Coma cluster was
estimated to be $T_{SZ} = 6.6 \pm 13$ keV, using purely the SZ
observations from the Coma cluster~\cite{hansencoma}.  Worse yet, the
normal connections between thermodynamical quantities (e.g. how
temperature is related to pressure and density) will have to be
reconsidered, since they were derived under the normal assumption of
extensive entropy. Clearly this may affect the cooling flow problem,
however, we leave a detailed study of this aspect for the future.
Since the derived temperature may differ from the true thermodynamic
temperature, then the M-T and L-T relations will be affected.  One
could imagine that the smaller structure will have a relatively larger
fraction of the cluster which should be described by non-extensive
thermodynamics, in which case the lower temperature clusters have a
true temperature which is lower than the observed one. This could push
the low-T tail of the L-T relation towards the theoretically expected
values.  Also the tests of DM structure will be
different~\cite{lewis,pointec} where e.g. a lower central temperature
(as exemplified above in NGC 4874) would lead to a shallower DM profile.

Let us finalize by reconsidering the spectral lines~\cite{silva98}. 
The temperature
determination from the 'exponential' tail is very accurate, but it
is always nice with an alternative consistency check. The temperature
can in principle be measured directly from the Doppler broadening
of the spectral lines~\cite{rybicki}, e.g. through the Doppler
width $\Delta \nu_D = \nu/c \sqrt{2 {\rm ln}2 /m\beta}$, where $\beta$ is the
inverse temperature. In the generalized case this becomes
\begin{equation}
\Delta \nu _D = \frac{\nu}{c} \sqrt{\frac{-2\alpha}{m\beta_q} \, {\rm ln}_q
\left( 1/2 \right) ^{1/q}} \, .
\end{equation}
This method requires somewhat better energy resolution that what we
have today. This method requires that the line shape is sufficiently
accurately determined that one can separate turbulent gas motion from
the microphysical distribution function.  Other methods of temperature
determination, e.g. through line ratios is more complicated to
calculate for the generalized statistics, but may be interesting to
consider given that we already have sufficient resolution to use
this~\cite{nevalainen}.

\section{Conclusions}
\label{sec:con}
We propose to use non-extensive thermo-statistics to analyse galaxy
clusters. We explain that there is growing theoretical indications
that this Tsallis statistics is the correct statistics to consider in
the central part of cosmological structures.

The use of such generalized thermo-statistics leads to a difference
between the observed and the true temperature as explained through
eqs.~(\ref{eq:alpha}, \ref{eq:beta}). We show that this difference may
be as big as a factor 5/2.  Thus, the observed low central cluster
temperature may be different from what we believe today,
both higher or lower.  In general the M-T and L-T relations will be
affected, as well as the central DM slope derived through hydrostatic
equilibrium may be either more shallow or steeper.

We show how a correct temperature determination can be done either
through the use of the quasi-exponential continuum spectrum, or
through the Doppler broadening of the spectral lines.

We have raised more questions than given answers, and it will be very
interesting in the near future to test our proposal on dynamically
young structures, such as ongoing mergers and shocks in galaxy
clusters, which are only kinetically equilibrated.

\section*{Acknowledgments}
It is a great pleasure to thank Monique Arnaud for kindly providing
the spectra from NGC 4874. The author thanks Ben Moore and Rocco
Piffaretti for discussions, and the Tomalla foundation for
financial support.

\end{document}